\documentclass[twocolumn,superscriptaddress,showpacs,preprintnumbers,amsmath,amssymb]{revtex4}

\usepackage{graphicx}
\usepackage{dcolumn}
\usepackage{bm}

\newcommand{\bra}[1]{{\left\langle #1 \right|}}
\newcommand{\ket}[1]{{\left| #1 \right\rangle}}
\newcommand{\inn}[2]{{\left\langle #1 | #2 \right\rangle}}

\begin{document}

\title{Convex-roof extended negativity as an
entanglement measure \\ for bipartite quantum systems}

\author{Soojoon Lee}\email{level@kias.re.kr}
\affiliation{
 School of Computational Sciences,
 Korea Institute for Advanced Study,
 Seoul 130-722, Korea
}
\author{Dong Pyo Chi}\email{dpchi@math.snu.kr}
\affiliation{
 School of Mathematical Sciences,
 Seoul National University,
 Seoul 151-742, Korea
}
\author{Sung Dahm Oh}\email{sdoh@sookmyung.ac.kr}
\affiliation{
 Department of Physics,
 Sookmyung Women's University,
 Seoul 140-742, Korea
}
\author{Jaewan Kim}\email{jaewan@kias.re.kr}
\affiliation{
 School of Computational Sciences,
 Korea Institute for Advanced Study,
 Seoul 130-722, Korea
}

\date{\today}

\begin{abstract}
We extend the concept of the negativity,
a good measure of entanglement for bipartite pure states,
to mixed states by means of the convex-roof extension.
We show that the measure does not increase
under local quantum operations and classical communication,
and derive explicit formulae for the entanglement measure
of isotropic states and Werner states,
applying the formalism presented
by Vollbrecht and Werner [Phys. Rev. A {\bf 64}, 062307 (2001)].
\end{abstract}

\maketitle

\section{Introduction}

Quantum information processing essentially depends on several quantum mechanical phenomena,
among which entanglement has been considered as one of the most crucial features.
There are two important problems for entanglement.
One is to find a method
to determine whether a given state in an arbitrary dimensional quantum system is separable or not,
and the other is to define the best measure quantifying an amount of entanglement of a given state.
In order to solve these problems,
various criteria for separability
and not a few measures of entanglement have been proposed in recent years
\cite{Peres,BDSW,Horodeckis1,Horodecki1,HW,VP,Wootters,
Horodeckis2,Horodeckis3,Horodecki,Wootters2,RBCHM,VidalW,VDC,RC}.
%
Although the perfect solutions for the problems have not yet been obtained,
quite a good criterion for separability,
called the {\em positive partial transposition} (PPT) criterion,
was suggested by Peres \cite{Peres} and Horodecki {\em et al.} \cite{Horodeckis1},
and an entanglement measure was naturally derived from the PPT criterion \cite{VidalW,ZHSL,Zyczkowski}.
The measure is called the {\em negativity} \cite{Negativity,VidalW},
and is defined by
\begin{equation}
\mathcal{N}(\rho)=\frac{\|\rho^{T_B}\|_1-1}{d-1},\label{eq:negativity}
\end{equation}
where $\rho^{T_B}$ is the partial transpose of a state $\rho$
in $d \otimes d'$ ($d\le d'$) quantum system
and $\|\cdot\|_1$ is the trace norm.

However, although the positivity of the partial transpose is a necessary and sufficient condition
for nondistillability in $2\otimes n$ quantum system \cite{Horodecki1,DCLB},
there exist entangled states with PPT in any bipartite system
except in $2\otimes 2$ and $2\otimes 3$ quantum systems \cite{Horodecki1,Horodeckis2},
that is, there exist entangled states whose negativity are not positive.
Such states are known as PPT bound entangled states,
which can be useful in a quasi-distillation process
called the activation of bound entanglement \cite{Horodeckis4,SST}.
Nevertheless, the negativity cannot distinguish the PPT bound entanglement from separability.
Hence, it is not sufficient for the negativity to be a good measure of entanglement
even in $2\otimes n$ quantum system.
In this paper, we present an extension of the negativity which can compensate for lack of the ability.

We now consider the negativity of pure states in $d\otimes d'$ $(d\le d')$ quantum system,
$\mathcal{H}_A\otimes \mathcal{H}_B$.
By the Schmidt decomposition theorem, a given pure state $\ket{\Psi}$ can be written as
\begin{equation}
\ket{\Psi}
=\sum_{j=0}^{d-1}\sqrt{\mu_j}\ket{a_j b_j}
=U_A\otimes U_B\ket{\Phi},\label{eq:Schmidt}
\end{equation}
where $\sqrt{\mu_j}$ are the Schmidt coefficients,
$U_A$ and $U_B$ are unitary operators defined by $U_A\ket{j}=\ket{a_j}$ and
$U_B\ket{j}=\ket{b_j}$ respectively, and
\begin{equation}
\ket{\Phi}=\sum_{j=0}^{d-1}\sqrt{\mu_j}\ket{j j}.
\label{eq:Phi}
\end{equation}
Let
$\ket{\Psi_{ij}^{\pm}}=\left(\ket{ij}\pm\ket{ji}\right)/{\sqrt{2}}$.
Then since the partial transpose of $\ket{\Phi}\bra{\Phi}$ is
\begin{eqnarray}
\ket{\Phi}\bra{\Phi}^{T_B}
&=& \sum_{k=0}^{d-1}\mu_k\ket{kk}\bra{kk}
+ \sum_{i<j}\sqrt{\mu_i \mu_j}\ket{\Psi_{ij}^+}\bra{\Psi_{ij}^+}\nonumber \\
&&+ \sum_{i<j}\left(-\sqrt{\mu_i
\mu_j}\right)\ket{\Psi_{ij}^-}\bra{\Psi_{ij}^-},\label{eq:T_B}
\end{eqnarray}
we have
\begin{eqnarray}
\mathcal{N}(\ket{\Psi}\bra{\Psi})
&=&\mathcal{N}(\ket{\Phi}\bra{\Phi})\nonumber\\
&=&\frac{2}{d-1}\sum_{i<j}\sqrt{\mu_i\mu_j}\nonumber\\
&\equiv&
\mathcal{N}_{\mathrm{p}}(\vec{\mu}),\label{eq:pure_negativity}
\end{eqnarray}
where $\vec{\mu}=(\sqrt{\mu_0},\sqrt{\mu_1},\ldots,\sqrt{\mu_{d-1}})$
is the Schmidt vector.
We note that $\mathcal{N}_{\mathrm{p}}(\vec{\mu})=0$ if and only if
$\ket{\Psi}$ is separable,
and that $\mathcal{N}_{\mathrm{p}}((1,1,\ldots,1)/\sqrt{d})=1$.
Thus $\mathcal{N}_{\mathrm{p}}$ can be a measure of entanglement
for bipartite pure states in any dimensional quantum system,
and can be extended to mixed states $\rho$ by means of the convex roof,
\begin{equation}
\mathcal{N}_{\mathrm{m}}(\rho) \equiv\min_{\sum_k p_k
\ket{\Psi_k}\bra{\Psi_k}=\rho} \sum_k p_k
\mathcal{N}_{\mathrm{p}}(\vec{\mu}_k), \label{eq:c_negativity}
\end{equation}
where $\vec{\mu}_k$ is the Schmidt vector of $\ket{\Psi_k}$.
The extended measure $\mathcal{N}_{\mathrm{m}}$ is called the
{\em convex-roof extended negativity} (CREN).
Then we can readily show that
$\mathcal{N}_{\mathrm{m}}(\rho)=0$ if and only if $\rho$ is separable.
This implies that
the CREN can recognize the difference between separability and PPT bound entanglement,
which cannot be done by the original negativity.
%
We can also show that $\mathcal{N}_{\mathrm{m}}$ is convex,
and that $\mathcal{N}_{\mathrm{m}}(\rho)\ge\mathcal{N}(\rho)$
by the convexity of the original negativity $\mathcal{N}$~\cite{VidalW}.
In $2\otimes 2$ quantum system,
it follows from a straightforward calculation
that the CREN $\mathcal{N}_{\mathrm{m}}$ is equivalent
to Wootters' concurrence \cite{Wootters,Wootters2}
since
$\mathcal{N}_{\mathrm{p}}(\vec{\mu}) =2\sqrt{\mu_0
\mu_1}=|\langle\Psi|\tilde{\Psi}\rangle|=C(\ket{\Psi})$,
where $\vec{\mu}$ is the Schmidt vector of $\ket{\Psi}$,
$|{\tilde{\Psi}}\rangle=\sigma_y\otimes\sigma_y\ket{\Psi^*}$,
and $C$ is Wootters' concurrence.
Therefore, the CREN can be considered as a generalized version of Wootters' concurrence
which is different from the \mbox{$I$-concurrence} \cite{RBCHM,RC}.

Even though
it is generally not so easy to calculate the value of the convex-roof extension of a pure-state measure,
we can simplify the computation of entanglement measures
for states that are invariant under a group of local symmetries
\cite{VidalW,RC,TV,VollbrechtW},
such as isotropic states \cite{Horodeckis3} and Werner states \cite{Werner}.

In this paper,
we derive explicit formulae for the CREN of isotropic states and Werner states,
exploiting the formalism presented by Vollbrecht and Werner \cite{VollbrechtW}.
This formalism originated from the method of Terhal and Vollbrecht \cite{TV},
who gave an exact formula for the entanglement of formation for isotropic states,
and a subsequent work by Rungta and Caves \cite{RC}
recently provided explicit expressions for the concurrence-based entanglement measures of isotropic states.
These computational results imply that the newly defined measure,
CREN, is an entanglement measure
not only to show the difference between separability and bound entanglement,
but also to be computed as well as other convex-roof extended measures of entanglement.

This paper is organized as follows.
In Sec.\ \ref{Sec:EM_LOCC}
we show that the CREN $\mathcal{N}_{\mathrm{m}}$ does not increase
under local quantum operations and classical communication (LOCC),
that is, it is an entanglement monotone.
In Sec.\ \ref{Sec:isotropic_Werner}
we provide explicit formulae for the CREN of isotropic states and Werner states
in $d\otimes d$ quantum system,
and use these formulae to compare the CREN with the original negativity.
Finally, in Sec.\ \ref{Sec:Conclusion} we summarize our results.

\section{Monotonicity of entanglement
under local quantum operations and classical communication}\label{Sec:EM_LOCC}

Monotonicity of entanglement under LOCC is considered as one of natural requirements
which good measures of entanglement must hold.
Vidal \cite{Vidal} gave a nice recipe for building entanglement monotones
in bipartite quantum system
by showing that the convex-roof extension of a pure-state measure $E$
satisfying the following two conditions is an entanglement monotone:
(i) For a reduced density matrix $\rho_A=\mathrm{tr}_B\ket{\Psi}\bra{\Psi}$ of a pure state $\ket{\Psi}$,
the function $f$ on the space of density matrices defined by $f(\rho_A)=E(\ket{\Psi})$
is invariant under unitary operations,
that is, for any unitary operator $U$
\begin{equation}
f(U\rho_A U^\dagger)=f(\rho_A).\label{eq:unitary_invariance}
\end{equation}
(ii) The function $f$ is concave,
that is, for any density matrices $\rho_1$, $\rho_2$, and any $\lambda\in [0,1]$,
\begin{equation}
f(\lambda\rho_1+(1-\lambda)\rho_2)\ge \lambda
f(\rho_1)+(1-\lambda)f(\rho_2). \label{eq:cancavity}
\end{equation}

In this section, we are going to show that the CREN is an entanglement monotone,
by verifying that $\mathcal{N}_{\mathrm{p}}$ satisfies the above conditions.
Let $f$ be the function defined by
$f(\rho)=\mathcal{N}_{\mathrm{p}}(\vec{\mu})$,
$\vec{\mu}$ being the vector with entries consisting of eigenvalues of $\rho$.
Then since
\begin{eqnarray}
\mathcal{N}_{\mathrm{p}}(\vec{\mu})&=&\frac{2}{d-1}\sum_{i<j}\sqrt{\mu_i\mu_j}\nonumber \\
&=&\frac{1}{d-1}\left(\sum_{i}\sqrt{\mu_i}\right)^2-\frac{1}{d-1},\label{eq:N_p_mu}
\end{eqnarray}
we have
\begin{equation}
f(\rho)=\frac{1}{d-1}\left(g(\rho)-1\right)\label{eq:f_g}
\end{equation}
where $g(\rho)=\left[\mathrm{tr}(\sqrt{\rho})\right]^2$.
Thus, in order for the CREN to be an entanglement monotone,
it suffices to show that the function $g$ is concave,
since $g$ clearly satisfies the invariance under unitary operations.
Let $\rho=\lambda\rho_1+(1-\lambda)\rho_2=\sum_j r_j\ket{\xi_j}\bra{\xi_j}$,
$\rho_1=\sum_k p_k\ket{\phi_k}\bra{\phi_k}$,
and $\rho_2=\sum_l q_l\ket{\psi_l}\bra{\psi_l}$ be eigenvalue decompositions.
Then we obtain
\begin{widetext}
\begin{eqnarray}
g(\rho)
&=&\left[\mathrm{tr}\left(\sqrt{\lambda\rho_1+(1-\lambda)\rho_2}\right)\right]^2 \nonumber\\
&=&\left(\sum_j\sqrt{\lambda\bra{\xi_j}\rho_1\ket{\xi_j}
+(1-\lambda)\bra{\xi_j}\rho_2\ket{\xi_j}}\right)^2\nonumber\\
&\ge&
\lambda\left(\sum_j\sqrt{\bra{\xi_j}\rho_1\ket{\xi_j}}\right)^2
+(1-\lambda)\left(\sum_j\sqrt{\bra{\xi_j}\rho_2\ket{\xi_j}}\right)^2\nonumber\\
&=& \lambda\left(\sum_j\sqrt{\sum_k
p_k|\inn{\xi_j}{\phi_k}|^2}\right)^2
+(1-\lambda)\left(\sum_j\sqrt{\sum_l q_l|\inn{\xi_j}{\psi_l}|^2}\right)^2\nonumber\\
&\ge&
\lambda\left(\sum_{j,k}|\inn{\xi_j}{\phi_k}|^2\sqrt{p_k}\right)^2
+(1-\lambda)\left(\sum_{j,l}|\inn{\xi_j}{\psi_l}|^2\sqrt{q_l}\right)^2\nonumber\\
&=& \lambda\left(\sum_{k}\sqrt{p_k}\right)^2
+(1-\lambda)\left(\sum_{l}\sqrt{q_l}\right)^2\nonumber\\
&=&\lambda g(\rho_1)+(1-\lambda)g(\rho_2),\label{eq:concave_ineq}
\end{eqnarray}
\end{widetext}
from the straightforward calculations and the concavity of the square root.
Hence we complete the proof that the CREN
$\mathcal{N}_{\mathrm{m}}$ is an entanglement monotone.

\section{Entanglement for isotropic states and Werner states}\label{Sec:isotropic_Werner}
In this section, we compute the CREN for isotropic states,
\begin{equation}
\rho_F=\frac{1-F}{d^2-1}\left(I-\ket{\Phi^+}\bra{\Phi^+}\right)
+F\ket{\Phi^+}\bra{\Phi^+}\label{eq:isotropic}
\end{equation}
with
\begin{equation}
\ket{\Phi^+}=\frac{1}{\sqrt{d}}\sum_{j=0}^{d-1}\ket{jj}
\label{eq:Phi_plus}
\end{equation}
and $F=\langle \Phi^+ |\rho_F|\Phi^+\rangle$,
and Werner states,
\begin{eqnarray}
\varrho_W&=& \frac{2(1-W)}{d(d+1)}
\left(\sum_{k=0}^{d-1}\ket{kk}\bra{kk}+\sum_{i<j}\ket{\Psi_{ij}^+}\bra{\Psi_{ij}^+}\right)
\nonumber \\
&&+\frac{2W}{d(d-1)}\sum_{i<j}\ket{\Psi_{ij}^-}\bra{\Psi_{ij}^-}\label{eq:Werner}
\end{eqnarray}
with
\begin{equation}
W=\mathrm{tr}\left(\varrho_W\sum_{i<j}\ket{\Psi_{ij}^-}\bra{\Psi_{ij}^-}\right),
\label{eq:Werner_W}
\end{equation}
employing the formulation suggested by Vollbrecht and Werner \cite{VollbrechtW}.
Before deriving explicit formulae for the CREN of those states,
we review the formulation of a convex-roof extended measure.
Let $S$ be the set of states in a given quantum system,
$P$ the set of all pure states in $S$, and
let $G$ be a compact group of symmetries acting on $S$ by $(U,\rho)\mapsto U\rho U^\dagger$.
We also assume that a pure-state measure $E$ defined on $P$ is invariant under any operation in $G$.
We now define the projection ${\bf P}:S\rightarrow S$ by ${\bf P}\rho=\int dU U\rho U^\dagger$
with the standard (normalized) Haar measure $dU$ on $G$,
and the function $\varepsilon$ on ${\bf P}S$ by
\begin{equation}
\varepsilon(\rho)=\min\{E(\ket{\Psi}):\ket{\Psi}\in P, {\bf
P}\ket{\Psi}\bra{\Psi}=\rho\}. \label{eq:epsilon}
\end{equation}
Then for $\rho\in{\bf P}S$, we have
\begin{equation}
\mathrm{co}E(\rho)=\mathrm{co}\varepsilon(\rho), \label{eq:co}
\end{equation}
where $\mathrm{co}f$ is the convex-roof extension of a function
$f$, in other words, it is the convex hull of $f$.
%
\subsection{Isotropic states}\label{SSec:isotropic}
The isotropic states $\rho_F$ in Eq.\ (\ref{eq:isotropic}) have the important property
that $\rho_F$ is separable if and only if $\rho_F$ has PPT if and only if $0\le F\le 1/d$ \cite{Horodeckis3,TV}.
Since $\mathcal{N}_{\mathrm{m}}(\rho_F)=0$ for $0\le F\le 1/d$,
we now assume that $F\ge 1/d$.
Let $\mathcal{T}_{\mathrm{iso}}$ be the $(U\otimes U^*)$-twirling operator defined by
$\mathcal{T}_{\mathrm{iso}}(\rho)=\int dU (U\otimes U^*)\rho(U\otimes U^*)^\dagger$,
where $dU$ denotes the standard Haar measure on the group of all $d\times d$ unitary operations.
Then the operator satisfies the following two properties,
$\mathcal{T}_{\mathrm{iso}}(\rho)=\rho_{F(\rho)}$ with $F(\rho)=\bra{\Phi^+}\rho\ket{\Phi^+}$,
and
$\mathcal{T}_{\mathrm{iso}}(\rho_F)=\rho_{F}$.
Applying $\mathcal{T}_{\mathrm{iso}}$ to the pure state $\ket{\Psi}$ of Eq.\ (\ref{eq:Schmidt}),
we have
\begin{equation}
\mathcal{T}_{\mathrm{iso}}(\ket{\Psi}\bra{\Psi})=\rho_{F(\ket{\Psi}\bra{\Psi})}
\equiv\rho_{F(\vec{\mu},V)},
\end{equation}
where $V=U_A^TU_B$ and
\begin{equation}
F(\vec{\mu},V)=|\inn{\Phi^+}{\Psi}|^2
=\frac{1}{d}\left|\sum_k\sqrt{\mu_k}V_{kk}\right|^2
\label{eq:F_mu_V}
\end{equation}
with $V_{ij}=\bra{i}V\ket{j}$.
Then the function $\varepsilon$ defined in Eq.\ (\ref{eq:epsilon}) becomes
\begin{equation}
\varepsilon(\rho_F)= \min_{\{\vec{\mu},V\}}\left\{\mathcal{N}_{\mathrm{p}}(\vec{\mu})
:\frac{1}{d}\left|\sum_k\sqrt{\mu_k}V_{kk}\right|^2=F\right\}.
\label{eq:epsilon_isotropic}
\end{equation}
For a unitary operator $V$,
we consider the function
\begin{equation}
N_V(F)\equiv
\min_{\vec{\mu}}\left\{\mathcal{N}_{\mathrm{p}}(\vec{\mu})
:\frac{1}{d}\left|\sum_k\sqrt{\mu_k}V_{kk}\right|^2=F\right\},
\label{eq:N_V}
\end{equation}
and let $\vec{\nu}$ be the Schmidt vector to provide the minimum
for the constraint in Eq.\ (\ref{eq:N_V}).
Then since $F(\vec{\nu},I)\ge F(\vec{\nu},V)$,
we obtain an inequality,
\begin{equation}
N_I(F(\vec{\nu},I))\le \mathcal{N}_{\mathrm{p}}(\vec{\nu})
=N_V(F(\vec{\nu},V)). \label{eq:ineq01}
\end{equation}
Thus since $N_I$ is monotone increasing [see Eq.\ (\ref{eq:N_I})]
and $V$ is arbitrary, we get
\begin{equation}
N_V(F)\ge N_I(F), \label{eq:ineq02}
\end{equation}
which implies that $\varepsilon(\rho_F)=N_I(F)$.
Therefore, it follows from Eq.\ (\ref{eq:co})
that
\begin{equation}
\mathcal{N}_{\mathrm{m}}(\rho_F)=\mathrm{co}N_I(F)=N_I(F),
\label{eq:main_isotropic}
\end{equation}
with
\begin{eqnarray}
N_I(F)&=&\min_{\vec{\mu}}\left\{\mathcal{N}_{\mathrm{p}}(\vec{\mu})
:\frac{1}{d}\left|\sum_k\sqrt{\mu_k}\right|^2=F\right\}\nonumber \\
&=&\min_{\vec{\mu}}\left\{\mathcal{N}_{\mathrm{p}}(\vec{\mu})
:1+(d-1)\mathcal{N}_{\mathrm{p}}(\vec{\mu})=Fd\right\}\nonumber \\
&=&\frac{Fd-1}{d-1}. \label{eq:N_I}
\end{eqnarray}
We note that the CREN is equivalent to the original negativity
for isotropic states, that is,
\begin{equation}
\mathcal{N}_{\mathrm{m}}(\rho_F)=\max\left\{\frac{Fd-1}{d-1},0\right\}=\mathcal{N}(\rho_F).
\label{eq:compare_isotropic}
\end{equation}
\subsection{Werner states}\label{SSec:Werner}
Let $\mathcal{T}_{\mathrm{wer}}(\rho)=\int dU (U\otimes U)\rho(U^\dagger\otimes U^\dagger)$
be the $(U\otimes U)$-twirling operator.
Then the operator and the Werner states $\varrho_W$ in Eq.\ (\ref{eq:Werner}) have the following properties
analogous to isotropic states,
$\mathcal{T}_{\mathrm{wer}}(\rho)=\varrho_{W(\rho)}$
with $W(\rho)=\mathrm{tr}(\rho \sum_{i<j}\ket{\Psi_{ij}^-}\bra{\Psi_{ij}^-})$,
$\mathcal{T}_{\mathrm{wer}}(\varrho_W)=\varrho_{W}$, and
$\varrho_W$ is separable if and only if $\varrho_W$ has PPT
if and only if $0\le W\le 1/2$ \cite{Werner,DCLB,VollbrechtW,Wootters2}.
We now assume that $W\ge 1/2$,
since $\mathcal{N}_{\mathrm{m}}(\varrho_W)=0$ for $0\le W\le 1/2$.
Applying $\mathcal{T}_{\mathrm{wer}}$ to the pure state $\ket{\Psi}$ of Eq.\ (\ref{eq:Schmidt}),
we also have
\begin{equation}
\mathcal{T}_{\mathrm{wer}}(\ket{\Psi}\bra{\Psi})=\varrho_{W(\ket{\Psi}\bra{\Psi})}
\equiv\varrho_{W(\vec{\mu},\Lambda)},
\end{equation}
where $\Lambda=U_A^\dagger U_B$ and
\begin{eqnarray}
W(\vec{\mu},\Lambda)
&=&\mathrm{tr}\left((I\otimes\Lambda)\sum_{i,j}
\sqrt{\mu_i\mu_j}\ket{ii}\bra{jj}(I\otimes\Lambda^\dagger)
{\bf\Psi^-}\right)\nonumber \\
&=&\frac{1}{2}\sum_{i<j}\left|\sqrt{\mu_i}\Lambda_{ji}-\sqrt{\mu_j}\Lambda_{ij}\right|^2
\label{eq:W_mu_Lambda}
\end{eqnarray}
with ${\bf\Psi^-}=\sum_{i<j}\ket{\Psi_{ij}^-}\bra{\Psi_{ij}^-}$
and $\Lambda_{ij}=\bra{i}\Lambda\ket{j}$.
Then the function $\varepsilon$ defined in Eq.\ (\ref{eq:epsilon}) becomes
\begin{widetext}
\begin{eqnarray}
\varepsilon(\varrho_W) &=&\min_{\{\vec{\mu},\Lambda\}}
\left\{\mathcal{N}_{\mathrm{p}}(\vec{\mu})
:\frac{1}{2}\sum_{i<j}\left|\sqrt{\mu_i}\Lambda_{ji}-\sqrt{\mu_j}\Lambda_{ij}\right|^2=W
\right\} \nonumber \\
&=&\min_{\{\vec{\mu},\Lambda\}}
\left\{\mathcal{N}_{\mathrm{p}}(\vec{\mu})
:1-\sum_i\mu_i|\Lambda_{ii}|^2-
2\sum_{i<j}\sqrt{\mu_i\mu_j}\Re(\Lambda_{ij}\Lambda_{ji}^*)=2W
\right\}, \label{eq:epsilon_Werner}
\end{eqnarray}
\end{widetext}
where $\Re(z)$ is the real part of $z$.
Since
\begin{eqnarray}
2W&=& 1-\sum_i\mu_i|\Lambda_{ii}|^2-
2\sum_{i<j}\sqrt{\mu_i\mu_j}\Re(\Lambda_{ij}\Lambda_{ji}^*)
\nonumber \\
&\le& 1+2\sum_{i<j}\sqrt{\mu_i\mu_j}\left|\Re(\Lambda_{ij}\Lambda_{ji}^*)\right|
\nonumber \\
&\le&
1+2\sum_{i<j}\sqrt{\mu_i\mu_j}\nonumber \\
&=&1+(d-1)\mathcal{N}_{\mathrm{p}}(\vec{\mu}), \label{eq:ineq03}
\end{eqnarray}
the following inequality holds under the constraints in
Eq.\ (\ref{eq:epsilon_Werner}),
\begin{equation}
\mathcal{N}_{\mathrm{p}}(\vec{\mu})\ge \frac{2W-1}{d-1}.
\label{eq:ineq04}
\end{equation}
We note that the equalities hold in
Eq.\ (\ref{eq:ineq03}) or Eq.\ (\ref{eq:ineq04})
if $\Lambda_{00}=0$, $\Lambda_{01}=1$,
$\Lambda_{10}=-1$, $\Lambda_{11}=0$, and
$\vec{\mu}=(\mu_0,\mu_1,0,\ldots,0)$.
Therefore, we obtain
\begin{equation}
\mathcal{N}_{\mathrm{m}}(\varrho_W)
=\mathrm{co}\varepsilon(\varrho_W)
=\varepsilon(\varrho_W)=\frac{2W-1}{d-1}.
\label{eq:main_Werner}
\end{equation}
We remark that
for $W\ge 1/2$
\begin{equation}
\mathcal{N}_{\mathrm{m}}(\varrho_W)=\frac{2W-1}{d-1}
\ge\frac{2}{d}\cdot\frac{2W-1}{d-1}=\mathcal{N}(\varrho_W),
\label{eq:comparison}
\end{equation}
and that in a higher dimensional quantum system the CREN is
not equal to the original negativity.
\section{Conclusions}\label{Sec:Conclusion}

In this paper, we introduced the concept of the CREN,
showed that the CREN is an entanglement monotone,
and derived explicit formulae
for the entanglement measure of isotropic states and Werner states.
As seen in the exact formulae for those states and the analysis on the formulae,
the mathematical expressions for the CREN are less complicated
than those of other convex-roof extended measures,
such as the entanglement of formation.
Furthermore,
although by the convexity of the CREN
one can readily show that
the CREN is equal to the original negativity for
two-parameter states in $2\otimes n$ quantum systems presented by Chi and Lee \cite{CL},
no analytical formula for the entanglement of formation
for $2\otimes n$ systems exists in the literature,
and it seems a very hard problem to develop one \cite{RFWerner}.
Hence, we consider that
the CREN is
a more effectively computable measure of entanglement than any other convex-roof extended measure.
Nevertheless,
the CREN can recognize the difference between separable states and PPT bound entangled states,
which cannot be done in the computation of the original negativity.
Therefore, in this sense we conclude that
the CREN is a good candidate for the entanglement measures
in bipartite quantum systems.

\newpage
\begin{acknowledgments}
The authors acknowledge the KIAS Quantum Information Group for
useful discussions, and
S.L. would like to thank Prof. R.F. Werner for very helpful advices and comments.
S.L. is supported by a KIAS Research Fund (No. 02-0140-001),
D.P.C. by a Korea Research Foundation Grant (KRF-2000-015-DP0031),
and
S.D.O. and J.K. by a Korea Research Foundation Grant (KRF-2002-070-C00029).
\end{acknowledgments}


\begin{thebibliography}{1}

\bibitem{Peres} A.~Peres,
Phys. Rev. Lett. {\bf 77}, 1413 (1996).

\bibitem{BDSW} C.H.~Bennett, D.P.~DiVincenzo, J.A.~Smolin, and W.K.~Wootters,
Phys. Rev. A {\bf 54}, 3824 (1996).

\bibitem{Horodeckis1} M.~Horodecki, P.~Horodecki, and R.~Horodecki,
Phys. Lett. A {\bf 223}, 1 (1996).

\bibitem{Horodecki1} P.~Horodecki,
Phys. Lett. A {\bf 232}, 333 (1997).

\bibitem{HW}
 S.~Hill and W.K.~Wootters,
 Phys. Rev. Lett., {\bf 78}, 5022 (1997).

\bibitem{VP} V.~Vedral and M.B.~Plenio,
Phys. Rev. A {\bf 57}, 1619 (1998).

\bibitem{Wootters} W.K.~Wootters,
Phys. Rev. Lett. {\bf 80}, 2245 (1998).

\bibitem{Horodeckis2} M.~Horodecki, P.~Horodecki, and R.~Horodecki,
Phys. Rev. Lett. {\bf 80}, 5239 (1998).

\bibitem{Horodeckis3} M.~Horodecki and P.~Horodecki,
Phys. Rev. A {\bf 59}, 4206 (1999).

\bibitem{Horodecki} M.~Horodecki,
Quantum Inf. Comput. {\bf 1}, 3 (2001).

\bibitem{Wootters2} W.K.~Wootters,
Quantum Inf. Comput. {\bf 1}, 27 (2001).

\bibitem{RBCHM} P. Rungta, V. Bu\v{z}ek, C.M. Caves, M. Hillery, and G.J. Milburn,
Phys. Rev. A {\bf 64}, 042315 (2001).

\bibitem{VidalW} G.~Vidal and R.F.~Werner,
Phys. Rev. A {\bf 65}, 032314 (2002).

\bibitem{VDC} G.~Vidal, W.~D\"{u}r, and J.I.~Cirac,
Phys. Rev. Lett. {\bf 89}, 027901 (2002).

\bibitem{RC} P.~Rungta and C.M.~Caves,
Phys. Rev. A {\bf 67}, 012307 (2003).

\bibitem{ZHSL} K.~Zyczkowski, P.~Horodecki, A.~Sanpera, and M.~Lewenstein,
Phys, Rev. A {\bf 58}, 883 (1998).

\bibitem{Zyczkowski} K.~Zyczkowski,
Phys, Rev. A {\bf 60}, 3496 (1999).

\bibitem{Negativity}
Vidal and Werner~\cite{VidalW} defined the negativity of a state $\rho$ as $(\|\rho^{T_B}\|_1-1)/2$,
which corresponds to the absolute value of the sum of negative eigenvalues of $\rho^{T_B}$,
and which vanishes for separable states.
For a pure maximally entangled state such as one of the Bell states,
this quantity is strictly less than one.
As shown in Eq.\ (\ref{eq:pure_negativity}),
the negativity of a pure state must be defined as in Eq.\ (\ref{eq:negativity}),
in order for any pure maximally entangled state in $d\otimes d'$ ($d\le d'$) quantum system
to have the negativity one.

\bibitem{DCLB} W.~D\"{u}r, J.I.~Cirac, M.~Lewenstein, and D.~Bru\ss,
Phys. Rev. A {\bf 61}, 062313 (2000).

\bibitem{Horodeckis4} M.~Horodecki, P.~Horodecki and R.~Horodecki,
Phys. Rev. Lett. {\bf 82}, 1056 (1999).

\bibitem{SST} P.W. Shor, J.A. Smolin, and B.M. Terhal,
Phys. Rev. Lett. {\bf 86}, 2681 (2001).

\bibitem{TV} B.M.~Terhal and K.G.H.~Vollbrecht,
Phys. Rev. Lett. {\bf 85}, 2625 (2000).

\bibitem{VollbrechtW} K.G.H.~Vollbrecht and R.F.~Werner,
Phys. Rev. A {\bf 64}, 062307 (2001).

\bibitem{Werner} R.F.~Werner,
Phys. Rev. A {\bf 40}, 4277 (1989).

\bibitem{Vidal} G.~Vidal,
J. Mod. Opt. {\bf 47}, 355 (2000).

\bibitem{CL} D.P.~Chi and S. Lee,
J. Phys. A accepted for publication, e-print quant-ph/0309073.

\bibitem{RFWerner} R.F.~Werner,
private communication (September, 2003).


\end{thebibliography}
\end{document}